\documentclass[12pt]{article}

\begin{document}

\begin{center}
{\bf Kalb-Ramond fields in the Petiau-Duffin-Kemmer formalism and scale invariance}\\
\vspace{5mm} S. I. Kruglov\\
\vspace{3mm} \textit{University of Toronto at Scarborough,\\
Physical and Environmental Sciences Department, \\
1265 Military Trail, Toronto, Ontario, Canada M1C 1A4} \\

\vspace{5mm}
\end{center}

\begin{abstract}

Kalb-Ramond equations for massive and massless particles are considered
in the framework of the Petiau-Duffin-Kemmer formalism. We obtain $10\times10$ matrices
of the relativistic wave equation of the first-order and solutions in the form of density matrix.
The canonical and Belinfante energy-momentum tensors are found. We investigate
the scale invariance and obtain the conserved dilatation current. It was demonstrated that
the conformal symmetry is broken even for massless fields.

\end{abstract}

\section{Introduction}

Antisymmetric tensor massless fields introduced in \cite{Ogievetsky}, \cite{Kalb} give dual description of the massless scalar fields (associated with zero helicity). Such fields appear naturally in the (super)string theories and D-branes \cite{Green}, \cite{Polchinski}, and supergravity \cite{Salam}. The anomaly cancelation in the superstring theories are due to antisymmetric tensor fields. These fields also play an important role in the noncommutative geometry
\cite{Seiberg}, and cosmology \cite{Vilenkin}, \cite{Davis}. The specific coupling term with
antisymmetric tensor fields can generate an effective topological mass for the massless vector fields \cite{Cremmer}, \cite{Allen}. Different aspects of the theory of antisymmetric tensor fields were considered in \cite{Aurilia}, \cite{Dvoeglazov}, \cite{Ruegg}.

We investigate the Kalb-Ramond fields for massive and massless particles
in the framework of the first-order relativistic wave equation (RWE)
(Petiau-Duffin-Kemmer \cite{Petiau}, \cite{Duffin}, \cite{Kemmer} formalism).

The paper is organized as follows. In Sec.2, we consider massive
and massless fields in the form of RWE of the first-order with two mass parameters.
Solutions of the wave equation for free fields are obtained in the form of projection matrices in Sec.3.
We obtain the canonical and symmetrical Belinfante energy momentum tensors and the dilatation
currents in Sec.4. It is shown that the scale invariance
is broken in the case of massive fields but in the massless case the modified
dilatation current is conserved. We make a conclusion in Sec.5.

The Euclidean metric is explored and the system of units $\hbar =c=1$ is used.

\section{The first-order RWE}

Let us introduce equations with two mass parameters for the
third rank antisymmetric tensor field
\[
\partial _\mu \psi _{[\mu \nu\alpha]}+m_1 \psi _{[\nu\alpha]}=0,
\]
\vspace{-8mm}
\begin{equation}
\label{1}
\end{equation}
\vspace{-8mm}
\[
\partial _\mu \psi _{[\nu\alpha]}+\partial _\alpha \psi _{[\mu\nu]} +
\partial _\nu \psi _{[\alpha\mu]} +m_2\psi _{[\mu \nu\alpha]}=0.
\]
that are the generalization of \cite{Ogievetsky},\cite{Kalb}. Putting $m_1=0$,
and renormalizing fields $m_2\psi _{[\mu \nu\alpha]}\rightarrow -\phi _{\mu \nu\alpha}$,
we arrive at the Kalb-Ramond massless field
equations. At $m_1=m_2=m$, one has the fields with the mass $m$. In the general case,
fields possess the mass $m=\sqrt{m_1m_2}$. The fields $\psi_{[\mu\nu]}$, $\psi_{[\mu\nu\alpha]}$
have the same dimension. Consider the wave function
\begin{equation}
\Psi (x)=\left\{ \psi _A(x)\right\} =\left(
\begin{array}{c}
\psi_{[\mu\nu]} (x)\\
\psi_{[\mu\nu\alpha]}(x)\\
\end{array}
\right) \hspace{0.5in}(A=[\mu\nu] , [\mu\nu\alpha]).\label{2}
\end{equation}
With the help of the elements of the entire matrix algebra $\varepsilon
^{A,B}$, with matrix elements and products \cite{Kruglov}:
$\left( \varepsilon ^{M,N}\right) _{AB}=\delta _{MA}\delta _{NB}$,
$\varepsilon ^{M,A}\varepsilon ^{B,N}=\delta
_{AB}\varepsilon ^{M,N}$, where $A,B,M,N=[\mu\nu],[\mu\nu\alpha]$,
and generalized Kronecker symbols
\[
\delta_{[\mu\nu][\alpha\beta]}=\delta_{\mu\alpha}\delta_{\nu\beta}- \delta_{\mu\beta}\delta_{\nu\alpha},
\]
\begin{equation}
\delta_{[\mu\rho\alpha][\nu\beta\gamma]}=\delta_{\mu\nu}\delta_{\rho\beta}\delta_{\alpha\gamma}+ \delta_{\mu\beta}\delta_{\rho\gamma}\delta_{\alpha\nu}+\delta_{\mu\gamma}\delta_{\rho\nu}\delta_{\alpha\beta}
\label{3}
\end{equation}
\[
-\delta_{\mu\beta}\delta_{\rho\nu}\delta_{\alpha\gamma}-\delta_{\mu\nu}\delta_{\rho\gamma}\delta_{\alpha\beta}
-\delta_{\mu\gamma}\delta_{\rho\beta}\delta_{\alpha\nu},
\]
the system of equations (1) becomes the first-order RWE
\begin{equation}
\left( \beta _\mu \partial _\mu + m_1P_1+ m_2P_2\right)
\Psi (x)=0,
 \label{4}
\end{equation}
where we have introduced $10\times 10$ matrices
\begin{equation}
\beta_\mu=\frac{1}{2}\left(\varepsilon ^{[\nu\alpha],[\mu\nu\alpha]}+ \varepsilon
^{[\mu\nu\alpha],[\nu\alpha]}\right),~~~P_1=\frac{1}{2}\varepsilon ^{[\mu\nu],[\mu\nu]},~~~
P_2=\frac{1}{6}\varepsilon ^{[\mu\nu\alpha],[\mu\nu\alpha]}. \label{5}
\end{equation}
A summation over all repeated indices is implied.
The Hermitian matrices $\beta_\mu^+=\beta_\mu$ obey the PDK algebra
\begin{equation}
\beta _\mu \beta _\nu \beta _\alpha +\beta _\alpha \beta _\nu
\beta _\mu =\delta _{\mu \nu }\beta _\alpha+\delta _{\alpha \nu
}\beta _\mu , \label{6}
\end{equation}
and (Hermitian) projection matrices $P_1$, $P_2$ satisfy the relations as follows:
\[
P_1^2=P_1,~~P_2^2=P_2,~~P_1+P_2=1,~~
P_1P_2=P_2P_1=0,
\]
\vspace{-8mm}
\begin{equation}
\label{7}
\end{equation}
\vspace{-8mm}
\[
\beta_\mu P_1+P_1\beta_\mu=\beta_\mu,~~ \beta_\mu
P_2+P_2\beta_\mu=\beta_\mu.
\]
The operator $P_1$ extracts the six-dimensional subspace
($\psi_{[\mu\nu]}$) of the wave function $\Psi$, and the projection
operator $P_2$ extracts the four-dimensional tensor subspace
corresponding to the $\psi_{[\mu\nu\alpha]}$. At $m_1=m_2=m$, Eq.(4),
becomes the PDK equation for massive fields
\begin{equation}
\left( \beta _\mu \partial _\mu + m\right) \Psi (x)=0 . \label{8}
\end{equation}
The massless PDK equation at $m_1=0$ reads
\begin{equation}
\left( \beta _\mu \partial _\mu + m_2P_2\right) \Psi (x)=0 .
\label{9}
\end{equation}
The generators of the Lorentz group are given by
\begin{equation}
J_{\mu\nu}=\beta_\mu\beta_\nu-\beta_\nu\beta_\mu=
\varepsilon^{[\gamma\nu],[\mu\gamma]}-\varepsilon^{[\gamma\mu],[\nu\gamma]}
+ \frac{1}{2}\left(\varepsilon^{[\mu\gamma\sigma],[\nu\gamma\sigma]}-
\varepsilon^{[\nu\gamma\sigma],[\mu\gamma\sigma]}\right),
 \label{10}
\end{equation}
and obey the commutation relations
\[
\left[J_{\rho\sigma},J_{\mu\nu}\right]=\delta_{\sigma\mu}J_{\rho\nu}+
\delta_{\rho\nu}J_{\sigma\mu}-\delta_{\rho\mu}J_{\sigma\nu}-\delta_{\sigma\nu}J_{\rho\mu},
\]
\vspace{-8mm}
\begin{equation}
\label{11}
\end{equation}
\vspace{-8mm}
\[
\left[\beta_{\lambda},J_{\mu\nu}\right]=\delta_{\lambda\mu}\beta_\nu
-\delta_{\lambda\nu}\beta_\mu,~~\left[P_1,J_{\mu\nu}\right]=
0,~~\left[P_2,J_{\mu\nu}\right]= 0 .
\]
The Hermitianizing matrix, $\eta $, is defined as
\begin{equation}
\eta =2\beta_4^2-1=\frac{1}{2}\varepsilon ^{[4bc],[4bc]}-\frac{1}{6}\varepsilon ^{[abc],[abc]}
+\frac{1}{2}\varepsilon^{[ab],[ab]}-\varepsilon ^{[4b],[4b]}, \label{12}
\end{equation}
and the Lorentz-invariant is $\overline{\Psi }\Psi =\Psi
^{+}\eta \Psi$ \cite{Ahieser}, where $\Psi ^{+}$ is the Hermitian-conjugated wave
function. The Hermitian matrix $\eta^+ =\eta $ obeys
the relations: $\eta \beta _m=-\beta _m\eta$ (m=1,2,3), $\eta
\beta _4=\beta _4\eta$. Using these relations, we
obtain the conjugated equation
\begin{equation}
\overline{\Psi}(x)\left( \beta _\mu \overleftarrow{\partial}_\mu -
m_1P_1- m_2P_2\right) =0 . \label{13}
\end{equation}
The Lagrangian can be chosen in the form
\[
\mathcal{L}=-\frac{1}{2}\overline{\Psi }(x)\left( \beta _\mu
\partial _\mu + m_1P_1 + m_2 P_2\right)\Psi (x)
\]
\vspace{-8mm}
\begin{equation}
\label{14}
\end{equation}
\vspace{-8mm}
\[
+ \frac{1}{2}\overline{\Psi }(x)\left(\beta _\mu
\overleftarrow{\partial _\mu} -m_1P_1- m_2 P_2\right) \Psi
(x).
\]
Eq.(4),(13) are obtained by varying the action corresponding to Lagrangian (14).
Let us consider for simplicity the neutral fields. With the help of Eq.(2),(12), one finds
the conjugated function
\begin{equation}
\overline{\Psi }(x)=\left( \psi_{[\mu\nu]}(x),
-\psi_{[\mu\nu\alpha]}(x)\right).
\label{15}
\end{equation}
The Lagrangian (14) with the help of Eq.(2),(5),(15) becomes
\begin{equation}
{\cal L}=\frac{1}{2}\psi_{[\mu\nu\alpha]}\partial
_\mu\psi_{[\nu\alpha]}-\frac{1}{2}\psi_{[\nu\alpha]}\partial _\mu\psi_{[\mu\nu\alpha]} -\frac{1}{2} m_1
\psi_{[\mu\nu]}^2 +
\frac{1}{6}m_2\psi_{[\mu\nu\alpha]}^2.
\label{16}
\end{equation}
Euler-Lagrange equations give the equations of motion (1). Lagrangians (14), and (16) become
zero for fields $\psi_A$ obeying Eq.(1) similar to the Dirac theory.

\section{Solutions to the matrix equation}

The solution to Eq.(8) for massive fields, in the form of the projection operator in the momentum space,
for the positive ($+p$) and negative ($-p$) energies, is given by (see, for example, \cite{Kruglov})
\begin{equation}
\Pi_\pm=\frac{\pm i\widehat{p}\left(\pm
i\widehat{p}-m\right)}{2m^2}, \label{17}
\end{equation}
where $\widehat{p}=\beta _\mu p _\mu$, and the four-momentum is
$p_\mu=(\textbf{p},ip_0)$ ($p^2= \textbf{p}^2-p_0^2$). Every column of the matrix (17) is the solution
to Eq.(8) in the momentum space, on-shell, when $p^2=-m^2$.
The Matrix $\Pi_\pm$ is the projection operator, $\Pi_\pm=\Pi_\pm^2$, and represents the density matrix.

For the case of massless particles, $m_1 =0$, the matrix of the equation (9),
in the momentum space (for the positive energy), is $\Lambda^{(0)}= i\widehat{p} + m_2 P_2$
and satisfies the minimal matrix equation for any $p^2$:
\begin{equation}
\Lambda^{(0)}\left(\Lambda^{(0)}-m_2\right)\left[\Lambda^{(0)}\left(\Lambda^{(0)}
-m_2\right)+p^2\right] =0.
\label{18}
\end{equation}
The solution to Eq.(9), in the form the density matrix (off-shell) is given by
\begin{equation}
\Pi^{(0)}=-\frac{1}{m_2p^2}\left(\Lambda^{(0)}-m_2\right)\left[\Lambda^{(0)}\left(\Lambda^{(0)}
-m_2\right)+p^2\right]=\frac{p_\mu p_\nu}{2p^2}\varepsilon^{[\mu\gamma],[\nu\gamma]}, \label{19}
\end{equation}
so that the $\Pi^{(0)}$ is the projection operator, $\Pi^{(0)2}=\Pi^{(0)}$.
It should be noted that Eq.(19) is valid only off-shell, $p^2\neq 0$. If $p^2=0$,
Eq.(18) becomes
\begin{equation}
\Lambda^{(0)2}\left(\Lambda^{(0)}-m_2\right)^2 =0,
\label{20}
\end{equation}
and zero eigenvalues of the operator $\Lambda^{(0)}$ are degenerated.
In this case it is impossible to obtain solutions in the
form of projection matrix \cite{Fedorov}. As Kalb-Ramond fields are longitudinal fields
(with the helicity h = 0), we do not need the spin operators. The density matrix (17) can be used for
different quantum calculations with Kalb-Ramond fields. These fields also give the dual description of
scalar fields (because the helicity is zero).

\section{The energy-momentum tensor and scale invariance}

From Lagrangian (14), using the general procedure \cite{Ahieser},
we obtain the canonical energy-momentum tensor
\begin{equation}
T^c_{\mu\nu}=\frac{1}{2}\left(\partial_\nu \overline{\Psi}
(x)\right)\beta_\mu \Psi (x)-\frac{1}{2} \overline{\Psi}
(x)\beta_\mu \partial_\nu\Psi (x). \label{21}
\end{equation}
We took into account that the Lagrangian (14) vanishes
for fields obeying the equations of motion. With the help of
Eq.(2),(5),(15), the canonical energy-momentum tensor (21) becomes
\begin{equation}
T^c_{\mu\nu}=\frac{1}{2}\left[\psi_{[\mu\beta\alpha]}\partial_\nu
\psi_{[\beta\alpha]}-\psi_{[\beta\alpha]}\partial_\nu \psi_{[\mu\beta\alpha]}\right].
\label{22}
\end{equation}
One can verify, with the aid of Eq.(1), that the energy-momentum
tensor (22) is conserved tensor, $\partial_\mu
T^c_{\mu\nu}=0$. The canonical energy-momentum tensor is not the
symmetric tensor, and its trace does not equal zero:
\begin{equation}
T_{\mu\mu}^{c}=\frac{1}{2}m_1\psi_{[\mu\nu]}^2-\frac{1}{6}m_2
\psi_{[\mu\nu\alpha]}^2. \label{23}
\end{equation}

To investigate the dilatation symmetry, we explore the method of
\cite{Coleman} adopted to first-order formalism \cite{Kruglov1}. The
canonical dilatation current is given by \cite{Coleman}
\begin{equation}
D_\mu^c=x_\alpha T_{\mu\alpha}^{c}+\Pi_\mu \textbf{d}\Psi,
\label{24}
\end{equation}
where
\begin{equation}
\Pi_\mu=\frac{{\partial\cal
L}}{\partial\left(\partial_\mu\Psi\right)} =-\overline{\Psi
}\beta_\mu, \label{25}
\end{equation}
and the matrix $\textbf{d}$ is defined by the field dimension. In the
case of Bose fields the $\textbf{d}$ is the unit matrix. From
Eq.(24),(25), one finds
\begin{equation}
\partial_\mu D_\mu^c=T_{\mu\mu}^{c}, \label{26}
\end{equation}
where we use the conservation of the current $
\partial_\mu j_\mu=\partial_\mu(i\overline{\Psi}\beta_\mu\Psi)=0$. It should be noted that for neutral fields the electric current vanishes, $j_\mu=0$. One can verify with the help of Eq.(2),(5),(15) that $\overline{\Psi}\beta_\mu\Psi=0$.
The relation (26) also follows from \cite{Coleman}
\[
\partial_\mu D_\mu^c=\Pi_\mu\left(\textbf{d}+1\right)\partial_\mu\Psi
 +\frac{\partial{\cal L}}{\partial\Psi}\textbf{d}\Psi
 +\overline{\Psi}\textbf{d}\frac{\partial{\cal L}}{\partial\overline{\Psi}}-4{\cal L}
\]
\vspace{-8mm}
\begin{equation}
\label{27}
\end{equation}
\vspace{-8mm}
\[
=\overline{\Psi
}\left(m_1P_1+m_2P_2\right)\Psi=\frac{1}{2}m_1\psi_{[\mu\nu]}^2-\frac{1}{6}m_2
\psi_{[\mu\nu\alpha]}^2.
\]
The dilatation symmetry is broken due to the presence of massive
parameters $m_1$ and $m_2$. For the massless fields, $m_1=0$, the
dilatation current $D_\mu^c$ is also not conserved, but we
will introduce new conserved current.

Let us consider the Belinfante tensor \cite{Coleman}
 \begin{equation}
T_{\mu\alpha}^{B}=T_{\mu\alpha}^{c}+
\partial_\beta X_{\beta\mu\alpha}, \label{28}
\end{equation}
where
\begin{equation}
X_{\beta\mu\alpha}=\frac{1}{2}\left[\Pi_\beta J_{\mu\alpha}\Psi-
\Pi_\mu J_{\beta\alpha}\Psi-\Pi_\alpha
J_{\beta\mu}\Psi\right]. \label{29}
\end{equation}
From Eq.(2),(5),(10),(15),(25), we find the tensor
$X_{\beta\mu\alpha}$:
\[
X_{\beta\mu\alpha}=\overline{\Psi}\beta_\beta\beta_\alpha\beta_\mu\Psi-
\delta_{\mu\alpha}\overline{\Psi}\beta_\beta\Psi
\]
\vspace{-8mm}
\begin{equation}
\label{30}
\end{equation}
\vspace{-8mm}
\[
=\frac{1}{2}\left(\delta_{\beta\alpha}
\psi_{[\mu\lambda\sigma]}\psi_{[\lambda\sigma]}-
\delta_{\alpha\mu}\psi_{[\beta\lambda\sigma]}\psi_{[\lambda\sigma]}\right)-2
\psi_{[\beta\lambda\mu]}\psi_{[\alpha\lambda]}.
\]
Using expressions (28),(30), and equations of motion,
one obtains the Belinfante energy-momentum tensor
\[
T_{\rho\nu}^{B}=-2m_1\psi_{[\rho\alpha]}\psi_{[\nu\alpha]}
+\delta_{\rho\nu}\left(\frac{1}{2}m_1\psi_{[\alpha\beta]}^2+\frac{1}{6}m_2   \psi_{[\mu\alpha\beta]}^2\right)
\]
\vspace{-8mm}
\begin{equation}
\label{31}
\end{equation}
\vspace{-8mm}
\[
+
\psi_{[\rho\alpha\beta]}\partial_\nu\psi_{[\alpha\beta]} -
2\psi_{[\mu\alpha\rho]}\partial_\mu\psi_{[\nu\alpha]}.
\]
From Eq.(31), we find the trace of the Belinfante energy-momentum tensor
\begin{equation}
T_{\mu\mu}^{B}=-\frac{1}{3}m_2\psi_{[\mu\alpha\beta]}^2.\label{32}
\end{equation}
The modified Belinfante dilatation current is defined as \cite{Coleman}
 \begin{equation}
D_\mu^B=x_\alpha T_{\mu\alpha}^{B}+V_\mu, \label{33}
\end{equation}
where the field-virial $V_\mu$ is
 \begin{equation}
V_\mu=\Pi_\mu \Psi-\Pi_\alpha
J_{\alpha\mu}\Psi=-\overline{\Psi}\beta_\mu\Psi+\overline{\Psi
}\beta_\alpha J_{\alpha\mu}\Psi. \label{34}
\end{equation}
With the help of (2),(5),(15), one obtains
\begin{equation}
V_\mu=-\frac{1}{2}\psi_{[\mu\alpha\beta]}\psi_{[\alpha\beta]},~~
\partial_\mu V_\mu=\frac{1}{2}m_1\psi_{[\mu\nu]}^2+\frac{1}{6}m_2
\psi_{[\mu\nu\alpha]}^2. \label{35}
\end{equation}
As a result, we find
\begin{equation}
\partial_\mu D_\mu^B=T_{\mu\mu}^{B}+\partial_\mu V_\mu=\frac{1}{2}m_1\psi_{[\mu\nu]}^2-\frac{1}{6}m_2
\psi_{[\mu\nu\alpha]}^2=\partial_\mu D_\mu^c.
\label{36}
\end{equation}
Thus, the divergence of the modified Belinfante dilatation current equals
the divergence of the canonical dilatation current. For the massive fields,
$m_1=m_2$, the dilatation symmetry is broken. In the case of massless
fields ($m_1=0$), the currents $D_\mu^c$, $D_\mu^B$ are not
conserved. We introduce the new conserved
current\footnote{I am grateful to Yu Nakayama for his remarks.}
\begin{equation}
D_\mu=D_\mu^B+V_\mu=x_\alpha T_{\alpha\mu}^{B}+2V_\mu, \label{37}
\end{equation}
and $\partial_\mu D_\mu=0$. The massless fields possess the dilatation symmetry
with the new dilatation current (37). However, the field-virial $V_\mu$ (35) is not the total divergence
of some local quantity $\sigma_{\alpha\beta}$, and, therefore, the conformal symmetry
is broken \cite{Coleman} even for the massless Kalb-Ramond fields.

\section{Conclusion}

We have considered the massive and massless Kalb-Ramond fields in the
PDK formalism. Solutions in the form of density matrix obtained
allow us to make calculations of quantum processes
with particles associated with Kalb-Ramond fields in the covariant form \cite{Fedorov}.
The canonical and symmetric Belinfante energy-momentum tensors found possess their nonzero traces. We
have investigated the scale invariance in the first-order formalism.
It was shown that the dilatation symmetry is broken for
massive fields but in the massless case the new dilatation
current is conserved. The conformal symmetry is broken
as for massive as for the massless Kalb-Ramond fields.

\end{document}